\documentclass[journal]{IEEEtran}

\usepackage{amssymb}
\usepackage{amsmath,amsthm}
\usepackage{amsthm}
\usepackage{epstopdf}
\usepackage[caption=false,font=footnotesize]{subfig}
\usepackage[pdftex]{graphicx}
\usepackage{multirow}
\usepackage{etoolbox}

\usepackage{xspace}
\newcommand{\fig}[1]     {Fig.~\ref{#1}}
\newcommand{\tab}[1]     {Table~\ref{#1}}

\newcommand{\secref}[1]  {Sec.~\ref{#1}}
\newcommand{\eqnref}[1]  {Eq.~\eqref{#1}}

\newcommand{\dEdx}       {\ensuremath{dE/dx}\xspace}
\newcommand{\ERecoil}    {\ensuremath{E_{recoil}}\xspace}
\newcommand{\LO}         {\ensuremath{L}\xspace}
\newcommand{\TTT}        {\ensuremath{S}\xspace}
\newcommand{\LOave}      {\ensuremath{\hat{\LO}}\xspace}
\newcommand{\TTTave}     {\ensuremath{\hat{\TTT}}\xspace}
\newcommand{\LOaverange} {\ensuremath{\hat{\LO}\pm\sigma}\xspace}

\newcommand{\ALO}        {\ensuremath{A_{\LO}}\xspace}
\newcommand{\ATTT}       {\ensuremath{A_{\TTT}}\xspace}
\newcommand{\sobs}       {\ensuremath{\sigma_{\mathrm{obs}}}\xspace}
\newcommand{\sstat}      {\ensuremath{\sigma_{\mathrm{stat}}}\xspace}
\newcommand{\sanis}      {\ensuremath{\sigma_{\mathrm{anis}}}\xspace}
\newcommand{\sother}     {\ensuremath{\sigma_{\mathrm{other}}}\xspace}
\DeclareMathOperator\erf{erf}

\begin{document}

\title{Investigating the anisotropic scintillation response in anthracene through neutron, gamma-ray, and muon measurements}

\author{Patricia~Schuster,~\IEEEmembership{Member,~IEEE,}
        Erik~Brubaker,~\IEEEmembership{Member,~IEEE}
\thanks{P. Schuster is with the Department of Nuclear Engineering, University of California, Berkeley, CA, USA e-mail: pfschus@berkeley.edu.}
\thanks{E. Brubaker is with Sandia National Laboratories, Livermore, CA,  USA.}
\thanks{Manuscript received October 30, 2015.}}

\markboth{IEEE Transactions on Nuclear Science}%
{Schuster \MakeLowercase{\textit{et al.}}: Investigation of the anisotropic scintillation response in anthracene}


\maketitle

\begin{abstract}
This paper reports a series of measurements that characterize the directional dependence of the scintillation response of crystalline anthracene to incident DT neutrons, DD neutrons, Cs-137 gamma rays, and, for the first time, cosmic ray muons. The neutron measurements give the amplitude and pulse shape dependence on the proton recoil direction over one hemisphere of the crystal, confirming and extending previous results in the literature. In similar measurements using incident gamma rays, no directional effect is evident, and any anisotropy with respect to the electron recoil direction is constrained to have a magnitude of less than a tenth of that present in the proton recoil events. Cosmic muons are measured at two directions, and no anisotropy is observed. This set of observations indicates that high \dEdx is necessary for an anisotropy to be present for a given type of scintillation event, which in turn could be used to discriminate among different hypotheses for the underlying causes of the anisotropy, which are not well understood.
\end{abstract}


\IEEEpeerreviewmaketitle

\section{Introduction}\label{Sec:Introduction}

\IEEEPARstart{O}{rganic} scintillator materials have long been used for radiation detection. They are particularly useful for their ability to detect both neutrons and gamma rays and to distinguish between them using pulse shape discrimination (PSD). Organic scintillators exist in plastic, liquid, and crystal forms. Compared to liquids, the solid plastic and crystal scintillators are easy to work with because they are subject to less thermal expansion and there is no risk of leaks. However, crystal scintillators are fragile, limited in size, and exhibit a directional variation~\cite{Knoll}. For these reasons, many users opt to use liquid and plastic materials.

There is renewed interest in organic crystal scintillators following a new growing method  that produces large crystals with excellent light output and neutron-gamma PSD~\cite{Carman201356}. Still, the directional variation remains as a limit to the performance. When a heavy charged particle deposits energy in an organic crystal scintillator, the light output and pulse shape may depend on the direction of the particle with respect to the crystal axes. This degrades the energy resolution and widens the distribution of pulse shapes produced in these materials when heavy charged particles (e.g. nuclear recoils from neutron interactions) interact at many angles in the crystal axes. 

For some applications, this serves as an obstacle, and these materials would serve better if one could correct for the directional dependence or synthesize new materials that eliminate it. Other applications exist in which the directional dependence could be exploited for a compact directional detection system. In this case, it may be preferred to use materials with a large directional dependence, or synthesize new materials with an enhanced directional dependence. For either application, a greater understanding of the mechanism that produces the directional dependence is important to correct for, exploit, enhance, or eliminate the effect.

In order to contribute to the understanding of these systems, a new characterization has been performed on the directional dependence of proton recoil events from neutron interactions in anthracene. These measurements serve to augment and confirm similar measurements made previously at a range of neutron energies.

Thus far, no directional dependence has been observed in electron recoil events produced by gamma-ray interactions, but no quantitative measurements have been published to demonstrate this. The mechanism that is responsible for the directional dependence in heavy charged particle interactions is not fully understood, but it has been hypothesized to result partly from preferred directions of molecular excitation transport in the crystal~\cite{Brooks197469}. Such a mechanism may produce a smaller but non-zero directional dependence for electron recoil events. Thus, a characterization has been performed on the directional response to electron recoil events produced by gamma-ray Compton scatter interactions in anthracene. These measurements will serve either to demonstrate that there is a directional dependence and measure its magnitude, or, if no effect is observed, to set an upper bound on its magnitude.

Heavy charged particle recoils differ from electron recoils in that they deposit their energy with a high stopping power \dEdx in a straight path. Electron recoils interact with relatively low \dEdx and do not travel along a straight path. Energetic cosmogenic muons are minimum ionizing particles that like electrons deposit energy with low \dEdx, but like protons travel in a straight path. Measurements of cosmic muons were therefore obtained to test for a directional dependence. The presence or absence of a directional dependence in cosmic muon interactions will probe whether a high stopping power \dEdx or straight trajectory is necessary for producing a directional dependence.

\section{General Techniques}\label{Sec:General Techniques}

\subsection{Excitations and Light Emission in Anthracene}\label{Sec:LightEmission} 
In order to understand the mechanism that produces the directional dependence in organic crystal scintillators, one must consider the molecular excitations produced by radiation interactions. A brief summary of relevant concepts is provided here; more detailed treatment is available elsewhere, e.g.~\cite{birks1964theory}. After radiation deposits energy in an organic scintillator, the system quickly relaxes into singlet (antiparallel spin electrons) and triplet (parallel spin electrons) molecular excitations of the delocalized $\pi$-orbitals. These excitations may undergo numerous kinetic processes including de-excitation by light emission. A de-excitation from the lowest singlet excited state to the ground state via light emission is known as fluorescence and occurs on the ns time scale. A de-excitation from the lowest triplet excited state to the ground state via light emission is known as phosphorescence and occurs on the $\mu$s-ms time scale, which is longer than the time scale of our measurement system so phosphorescence is essentially unobserved. 

Triplet energy may be observed on a shorter time scale through another kinetic process known as triplet-triplet annihilation~\cite{Azumi1963}. In this process, two triplet states in close proximity interact and annihilate into one singlet excited state and one singlet ground state. The singlet excited state may then de-excite by fluorescence on the ns time scale. This emission is known as delayed fluorescence, as the time of the light emission is determined by the time required for the two triplet states to travel through the material and annihilate. A higher rate of triplet-triplet annihilation will increase the amount of light in the delayed regions of the pulse. Thus, the amount of delayed fluorescence produced by triplet-triplet annihilation will depend on the density of triplet excitations in the material and their mobility.

Singlet excitations are also subject to interactive processes that affect their light emission. One such process is singlet ionization quenching, in which two singlet states interact, leaving one in ground state and the other in a super-excited singlet state. The super-excited singlet may then de-excite by light emission, halving the amount of light that could have been produced by the original two singlet excitations. Thus, the amount of prompt light produced by singlet fluorescence from the initial singlet population will be less for a system in which singlet excitations are produced at a higher density, increasing the fraction of delayed light.

The effects of these interactive processes are observed in the differences of the pulse shapes produced by neutron and gamma-ray events. Neutron interactions produce nuclear recoils in the material, which deposit their energy with much higher \dEdx than the electron recoils produced by gamma-ray interactions. The higher \dEdx produces higher excitation densities, more triplet-triplet annihilation, and more singlet quenching. Thus, there is relatively more delayed light and relatively less prompt light in the signal produced by neutron interactions. Not only is the time distribution of light emitted different, but the total amount of light emitted per energy deposited is different for neutron and gamma-ray events. The difference in the pulse shape allows for events to be identified as neutron or gamma-ray events through PSD. 

In crystals, the excitation density may depend on the direction of the recoil particle. Additionally, the rates at which kinetic processes occur may have a directional dependence. Either of these effects could contribute to the anisotropy observed in the scintillation output.

\subsection{Characteristics of Anthracene Crystal} 
Anthracene was chosen for this study because it has the largest magnitude of pulse shape anisotropy of the organic crystal scintillators that have been measured in the past~\cite{Brooks197469}.
If there is a small anisotropy in the gamma-ray response, it is more likely that a measurement system with a given sensitivity level could observe it in anthracene than in other materials because the effect is larger in anthracene. 

The anthracene crystal used in these measurements is an older sample with unknown history and considerable wear. Several minor cracks are visible within the crystal and the surface has been polished numerous times. The crystal was produced using a melt growth technique, so it is unlikely to be to a perfect monocrystal. The sample is a cylinder approximately 0.75" tall and 1" diameter. 

\begin{figure}[!t]
\centering
\subfloat[Cartoon of 3D vector direction in top hemisphere represented in spherical coordinates with ${0^\circ<\theta<90^\circ}$, ${0^\circ<\phi<360^\circ}$.]{\includegraphics[trim={3cm 2cm 2cm 2cm},clip,width=2.5in]{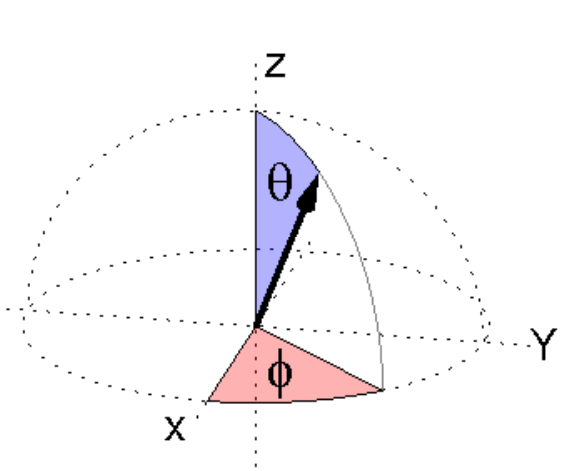}%
\label{Fig:SphericalCoordinatesCartoon}}
\hfil
\subfloat[2D representation of the top hemisphere with $\phi$ increasing counter-clockwise and $\theta$ increasing radially outward as $r=\sqrt{1-\cos\theta}$.]{\includegraphics[trim={2cm 0.6cm 2cm 0.3cm},clip,width=2.5in]{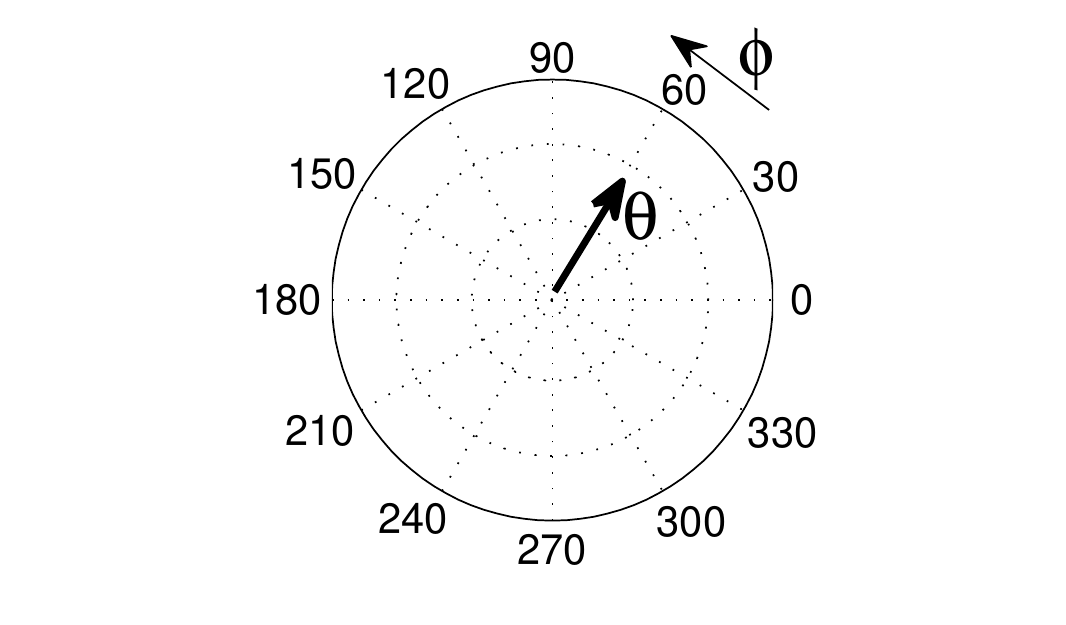}%
\label{Fig:SphericalCoordinates2D}}
\caption{2D and 3D visualization of directions in spherical coordinates.}
\label{Fig:SphericalCoordinatesDiagram}
\end{figure}

The crystal axis directions within the sample are unknown, so an arbitrary set of axes has been established for these measurements. The direction of the interacting particle will be described in spherical coordinates using $\theta$ and $\phi$. As shown in \fig{Fig:SphericalCoordinatesCartoon}, $\theta$ represents the angle between the direction and the positive $z$-axis, and $\phi$ represents the angle between the positive $x$-axis and the projection of the direction on the $xy$-plane. The arbitrary set of axes was maintained for all measurements so that the ($\theta$,$\phi$) coordinates in all sets of measurements are consistent.

Because an interacting particle produces the same excitations in the forward and backward directions, only one hemisphere worth of interaction directions must be measured. The directions in the top hemisphere of space can be represented in two dimensions as shown in \fig{Fig:SphericalCoordinates2D}.

\subsection{Equipment and Data Acquisition} 
The anthracene crystal was wrapped in teflon tape and mounted to the face of a 2" Hamamatsu H1949-50 photomultiplier tube (PMT) assembly using V-788 optical grease. A plastic sleeve was placed over the crystal and wrapped in black tape to block out external light. The coupling of the crystal and PMT were fixed for all measurements to control the light collection efficiency.

Events were recorded using a Struck SIS3350 500 MHz 12-bit digitizer. The high voltage, gain, and offset were adjusted so that the raw baseline of the negatively polarized pulse, calculated as the average of the first 85 samples, was approximately 3965, and the largest amplitude events used about 80\% of the dynamic range. For each event, 384 samples were recorded in digitizer channel units. Each raw pulse was subtracted from its baseline to produce the baseline-subtracted pulse. 
Triggering was performed with a constant fraction discriminator set low with respect to events used in the analysis.

\subsection{Calculating Light Output and Quantifying Pulse Shape}\label{Subsec:Calculating Light Output and Quantifying Pulse Shape} 
Each event in the detector produces a pulse with samples of amplitude $x_i$ in baseline subtracted digitizer channel units, where $i$ is the sample measured. The light output \LO is calculated as the sum of the baseline subtracted pulse multiplied by a calibration factor $C$ as shown in \eqnref{Eqn:LightOutput}.

\begin{equation}\label{Eqn:LightOutput}
	L = C \sum_{i=1}^{i=384} x_i
\end{equation}

The calibration factor $C$ is determined using a Cs-137 source producing mon\-o\-en\-er\-get\-ic gamma rays, and serves to convert the light output from summed digitizer channel units to keV-electron-equivalent (keVee). These units express the light output in terms of the energy that an electron would deposit in order to produce that number of optical photons.

In order to quantify the pulse shape, the pulse shape value \TTT is calculated as the fraction of light in a defined delayed region of the pulse as shown in \eqnref{Eqn:S}. 

\begin{equation}\label{Eqn:S}
	S = \frac{\sum_{i_2}^{i_3} x_i}{\sum_{i_1}^{i_3} x_i}
\end{equation}

The samples $i_1$, $i_2$, and $i_3$ define the beginning of the pulse, the beginning of the delayed region, and the end of the pulse, respectively. They are calculated as $i_1 = i_P - 10$, $i_2 = i_P + \Delta_1$, and $i_3 = i_P + \Delta_2$, where $i_P$ corresponds to the peak of the baseline subtracted pulse after a smoothing filter is applied with a smoothing span of 11 samples. The smoothing accounts for the jitter from fluctuations in photostatistics in order to pick out a consistent feature. $\Delta_1$ and $\Delta_2$ are selected via an iterative process to maximize separation between the distribution of \TTT values calculated for neutron and gamma-ray events. For this analysis on an anthracene detector using a digitizer measuring a sample every 2 ns, $\Delta_1 = 60$ samples and $\Delta_2 = 160$ samples. 

\section{Neutron Measurements}\label{Sec:Neutron Measurements}

\subsection{Purpose of Neutron Measurements} 
In order to confirm the directional dependence that has been documented in anthracene~\cite{Brooks197469,OliverKnoll1968,Tsukada1962286}, proton recoil events from neutron interactions have been measured. The directional dependence is characterized by measuring the expected expected light output \LOave and pulse shape \TTTave as a function of the recoil direction for protons of a fixed energy. The measurements presented in this paper have been made with digital pulse acquisition and processing, allowing for detailed offline analysis.

\subsection{Neutron Interactions in Anthracene} 
One interaction between a neutron and anthracene that produces measureable signal is an elastic scatter of a neutron on a $H$ nucleus, producing a proton recoil in the material. The proton recoil travels with energy \ERecoil~$=E_n \cos^2\alpha$, where $\alpha$ is the angle between the initial direction of the neutron and the proton recoil path. A proton that is scattered in the forward direction will travel with the full energy \ERecoil $=E_n$. This is the only proton recoil energy that corresponds to a unique direction, as a proton recoil that travels at a non-zero angle may be anywhere on the surface of a cone defined by the half-angle  $\alpha$ about the incident neutron direction. 

As the proton recoil travels through the anthracene crystal, it deposits its energy in a quasi-straight path with a relatively large \dEdx compared to electron recoils or cosmic muons. 

\subsection{Experimental Setup}\label{Sec:NeutronExperimentalSetup} 
In order to characterize the response of anthracene to proton recoil events at different directions within the crystal, the energy and direction of the proton recoil must be known. This was accomplished by selecting full energy proton recoil events produced by monoenergetic neutrons, fixing the proton recoil energy as $\ERecoil\approx E_n$ and the proton recoil direction as that of the incident neutron. In order to change the direction of the proton recoil in the crystal axes, the anthracene crystal was rotated to change its orientation with respect to the incident neutron direction.

\begin{figure}
	\centering
	\includegraphics[trim={0cm 0cm 0cm 0cm},clip,width=2in]{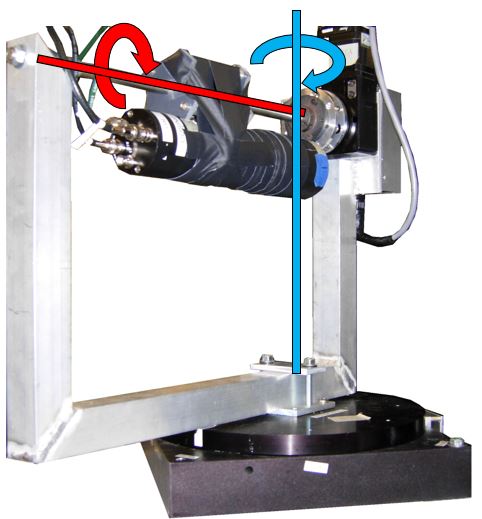}
	\caption{Photo of anthracene detector on rotational stage used in neutron and gamma-ray measurements showing the two motor-driven axes of rotation around 1) (blue) the vertical axis and 2) (red) the arm axis.
	\label{Fig:RotationalStagePhoto}
	}
\end{figure}

A Thermo Electric MP 320 neutron generator was used to produce neutrons via a DD (D+D$\rightarrow^3$He+$n$; $E_n$=2.5 MeV) or DT (D+T$\rightarrow^4$He+$n$; E$_n$=14.1 MeV) reaction. The anthracene detector was mounted to a rotational stage, shown in \fig{Fig:RotationalStagePhoto}, that is capable of positioning the detector at any angle in $4\pi$ with respect to the incident neutron direction. The stage has two motor-driven axes of rotation: 1) The circular turn table on which the support is mounted can rotate $360^\circ$ around the vertical axis, and 2) the metal arm on which the detector is mounted can rotate $360^\circ$ on its axis. For a given measurement, the incident neutron direction is calculated with respect to the crystal axes given the rotation angles of the two stage axes, the position of the generator, and the slight offset between the detector and the intersection of the stage axes. The anthracene crystal was approximately 60" away from the neutron generator, controlling the incident angle of the neutrons on the detector within 2$^\circ$.

For the DT measurements, 76 proton recoil directions were chosen to measure evenly across a hemisphere worth of directions. For the DD measurements, which are limited by lower flux produced by the neutron generator, 34 evenly distributed proton recoil directions were chosen.

\subsection{Data Analysis and Results}\label{Sec:NeutronDataAnalysis}
For each measurement, the following steps were taken to calculate \LOave and \TTTave, the expected \LO and \TTT values for a full-energy proton recoil at the angle of interest.

\begin{figure}
	\centering
	\includegraphics[scale=.8]{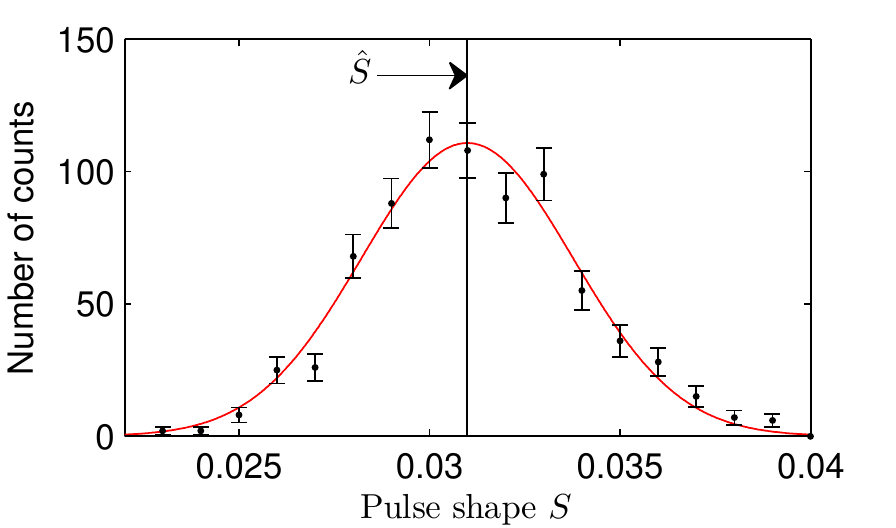}
	\caption{\TTT distribution for neutron events with \LO within \LOaverange produced by a DT neutron generator incident on anthracene at $(\theta,\phi)=(8.3^\circ,131.7^\circ)$. Black points are experimental data, the red line is the applied Gaussian fit function.
	\label{Fig:TTTGaussianFit}
	}
\end{figure}

\begin{enumerate}
\item Neutron selection: An \LO vs.\ \TTT distribution was produced, as shown in \fig{Fig:PSDPlot2D}. In this figure, the upper band with higher \TTT values is populated primarily by neutron events, and the lower band with lower \TTT values is populated primarily by gamma-ray events. The red lines indicate the light output threshold at 3000~keVee and cutoff for separating the neutron and gamma-ray events. Events above the red lines are selected for further analysis. 

\begin{figure}
	\centering
	\includegraphics[scale=.6]{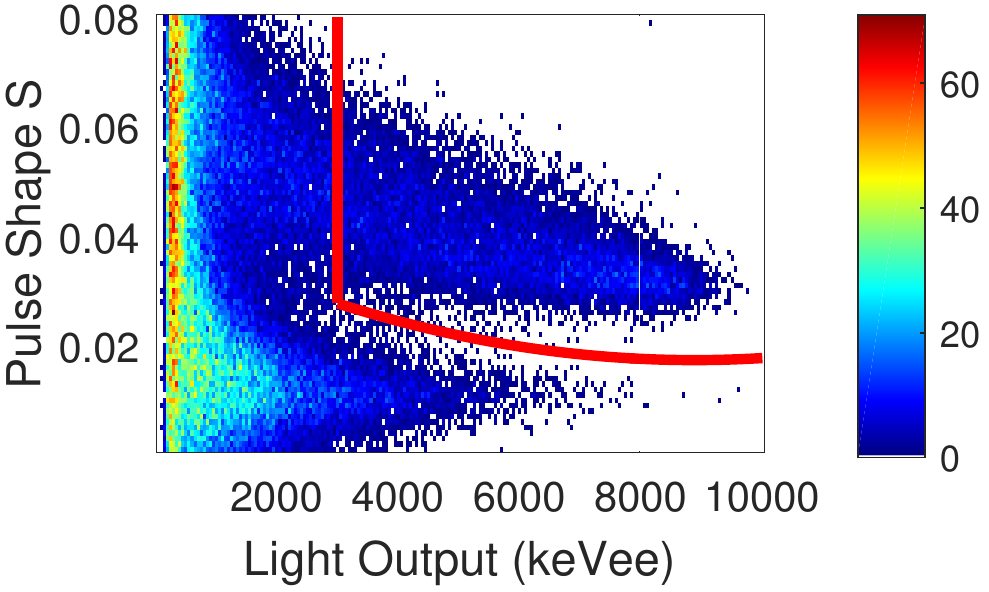}
	\caption{Density plot of \LO vs.\ \TTT for the mixed radiation field produced by a DT neutron generator incident on anthracene at $(\theta,\phi)=(8.3^\circ,131.7^\circ)$. The red lines are drawn to show the cutoff point for selecting neutron events above 3000 keVee.
	\label{Fig:PSDPlot2D}
	}
\end{figure}

\item Neutron light output spectrum fit: To calculate the light output at the spectrum endpoint, the energy spectrum of the selected events is fitted to the following function:
\begin{equation}\label{Eqn:LOFitFxn}
	f(L)=\frac{mL+b}{2}\left[1-\erf\left(\frac{L-\hat{L}}{\sigma\sqrt{2}}\right)\right]-\frac{m\sigma}{\sqrt{2\pi}}e^{\frac{-(L-\hat{L})^2}{2\sigma^2}}
\end{equation}
This function represents a sloped distribution with a hard cutoff convolved with a Gaussian resolution function. The value \LOave is the expected light output from a full energy proton recoil event, and $\sigma$ is the fitted resolution of the detector. A light output spectrum, along with its best-fit function, is shown in \fig{Fig:LOSpectrum}.

\begin{figure}
	\centering
	\includegraphics[scale=.8]{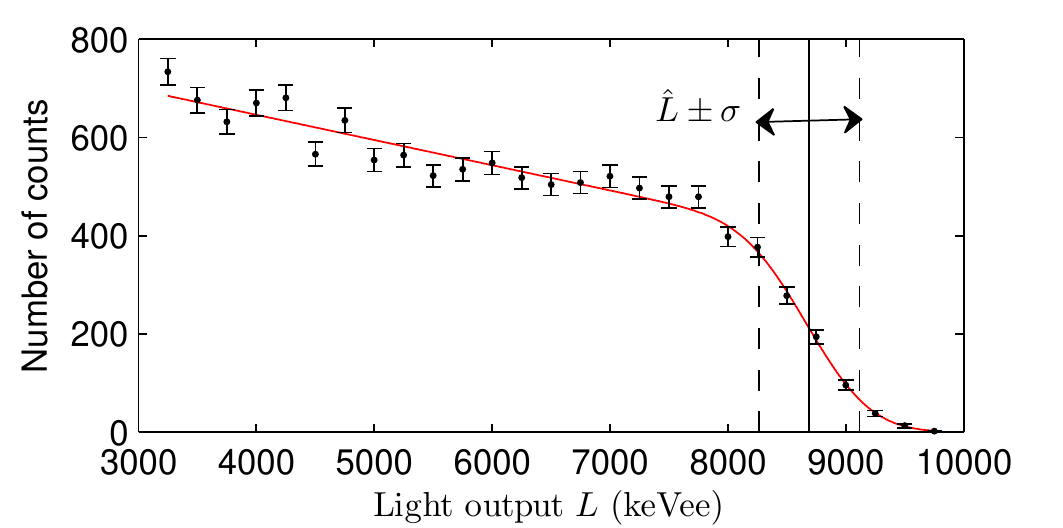}
	\caption{Light output spectrum for neutron events above 3000 keVee produced by a DT neutron generator incident on anthracene at $(\theta,\phi)=(8.3^\circ,131.7^\circ)$. Black points are experimental data, the red line is the applied fit function, and the range of full energy interactions \LOaverange is indicated.
	\label{Fig:LOSpectrum}
	}
\end{figure}

\item Full energy event selection: Events within the range \LOaverange are selected as full-energy proton recoils. For these measurements, this selection widens the range of proton recoil directions to events within $7^\circ$ of the forward direction.

\item Pulse shape distribution fit: A distribution of the \TTT value for full energy proton recoil events is produced. A Gaussian fit is applied to this distribution to calculate the centroid tail-to-total value \TTTave as shown in \fig{Fig:TTTGaussianFit}.
\end{enumerate}

\begin{figure}[!t]
\centering
\subfloat[Light output \LOave (keVee).]{\includegraphics[trim={3.1cm .7cm 1.5cm 0.4cm},clip,width=2.5in]{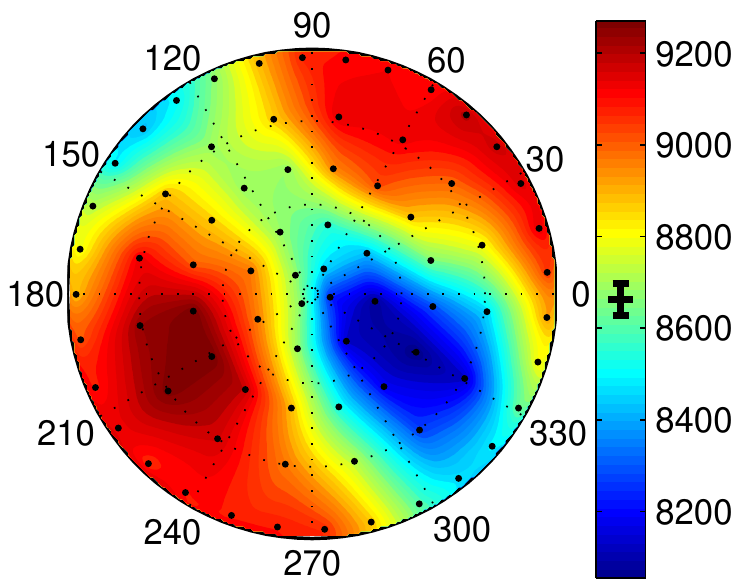}%
\label{Fig:2015_02_12_Anth_DT_LO_keVee_PolarPlotContour}}
\hfil
\subfloat[Pulse shape value \TTTave.]{\includegraphics[trim={3.1cm .7cm 1.5cm 0.4cm},clip,width=2.5in]{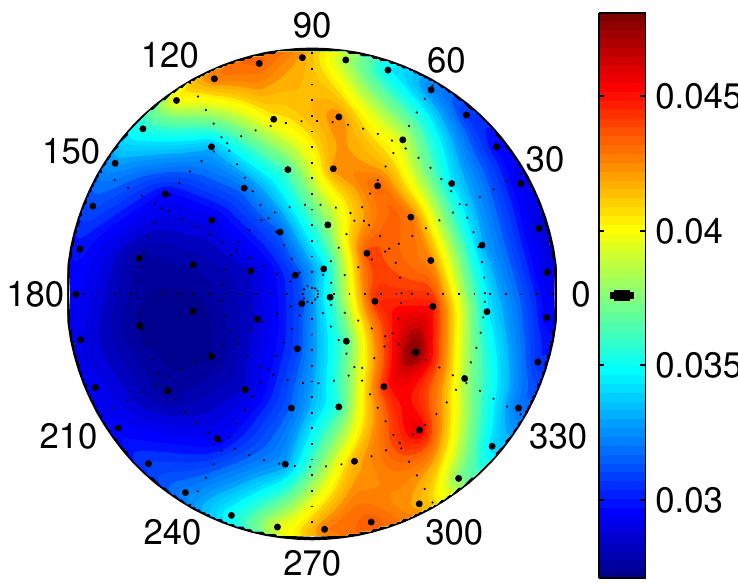}%
\label{Fig:2015_02_12_Anth_DT_TTT_PolarPlotContour}}
\caption{Response of anthracene crystal at various recoil directions to 14.1 MeV protons. Black points indicate measurements and the colors represent a smooth interpolation between measurements. Length of vertical black bar on colorbar indicates the statistical uncertainty.}
\label{Fig:2015_02_12_Anth_DT}
\end{figure}

For each measurement at a unique proton recoil direction, the \LOave and \TTTave values for full energy proton recoils are calculated. \fig{Fig:2015_02_12_Anth_DT} shows the \LOave and \TTTave values produced by 14.1 MeV proton recoils produced by a DT neutron generator at 76 directions in anthracene. The distributions show smooth transitions between maximum and minimum regions. 

\begin{figure}[!t]
\centering
\subfloat[Light output \LOave (keVee).]{\includegraphics[trim={3.1cm .7cm 1.5cm 0.4cm},clip,width=2.5in]{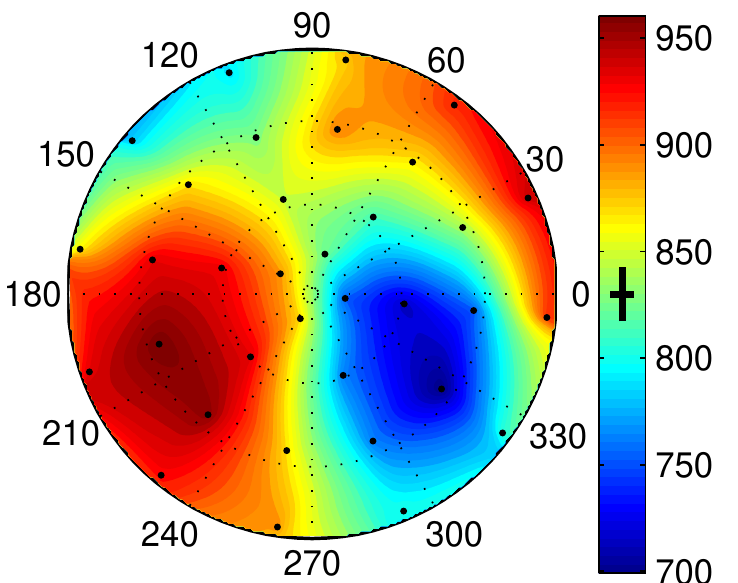}%
\label{Fig:2015_02_23_Anth_DD_LO_keVee_PolarPlotContour}}
\hfil
\subfloat[Pulse shape value \TTTave.]{\includegraphics[trim={3.1cm .7cm 1.5cm 0.4cm},clip,width=2.5in]{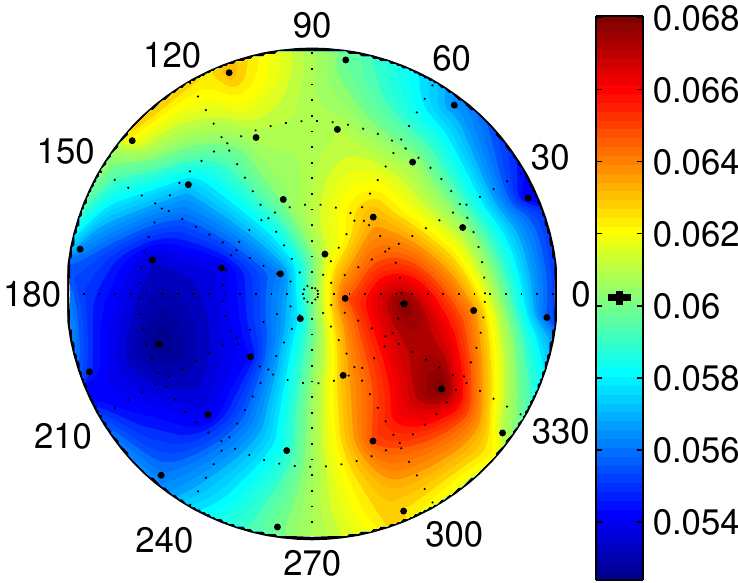}%
\label{Fig:2015_02_23_Anth_DD_TTT_PolarPlotContour}}
\caption{Response of anthracene crystal at various recoil directions to 2.5 MeV protons. Black points indicate measurements and the colors represent a smooth interpolation between measurements. Length of vertical black bar on colorbar indicates the statistical uncertainty.}
\label{Fig:2015_02_23_Anth_DD}
\end{figure}

\fig{Fig:2015_02_23_Anth_DD} shows the same distributions for 2.5 MeV proton recoils produced by a DD neutron generator at 34 directions in anthracene. Although there is less resolution in these distributions due to fewer measurements, the features are consistent with \fig{Fig:2015_02_12_Anth_DT}. 

To quantify the magnitude of the anisotropy at each energy, the ratio between the maximum and minimum observed values is calculated:
\begin{align*}
\ALO&=\frac{\LOave_{\mathrm{max}}}{\LOave_{\mathrm{min}}}&
\ATTT&=\frac{\TTTave_{\mathrm{max}}}{\TTTave_{\mathrm{min}}}
\end{align*}
These ratios for 14.1~MeV and 2.5~MeV proton recoil events in anthracene are shown in \tab{Table:AnthNeutronMagnitudeChange}. The errors are propagated from the errors in the calculation of \LOave and \TTTave as found by the fit function. These measurements demonstrate that the magnitude of the light output anisotropy is greater at lower proton recoil energies, while the magnitude of the pulse shape anisotropy is greater at higher proton recoil energies.

\begin{table}[ht]
\centering
\caption{Magnitude of Anisotropy Measured in Expected Light Output \LOave and Pulse Shape Value \TTTave Produced by Proton Recoil Events in Anthracene}
\label{Table:AnthNeutronMagnitudeChange}
\begin{tabular}{c|c|c}
    \ERecoil & 14.1 MeV          & 2.5 MeV \\
    \hline
    \ALO     & 1.155 $\pm$ 0.006 & 1.383 $\pm$ 0.023 \\
    \ATTT    & 1.798 $\pm$ 0.006 & 1.307 $\pm$ 0.005
\end{tabular}
\end{table}

\begin{figure}
	\centering
	\includegraphics[width=2.5in]{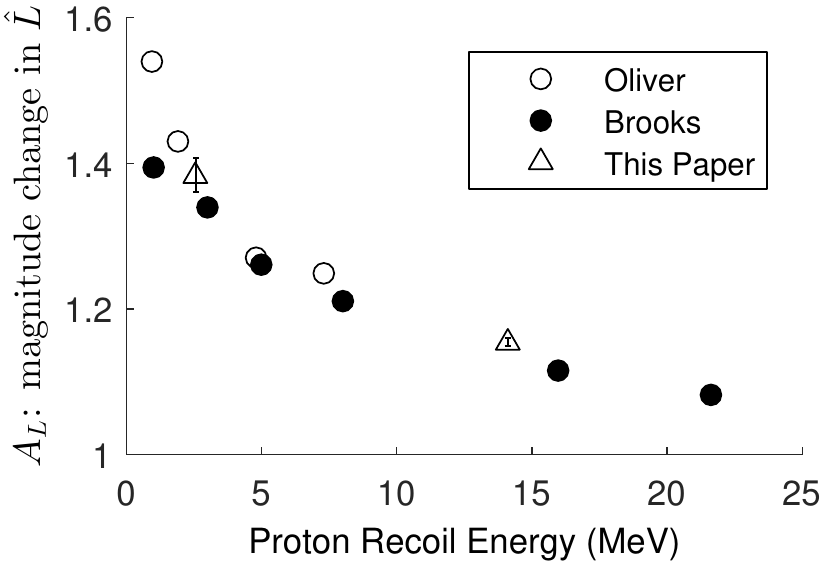}
	\caption{Magnitude of change in light output produced by proton recoil events in anthracene as a function of proton energy~\cite{Brooks197469,OliverKnoll1968}.
	\label{Fig:CompareOliverBrooks}
	}
\end{figure}

These measurements of the magnitude of change in the light output agree with past measurements made by others, as shown in \fig{Fig:CompareOliverBrooks}. All measurements are consistent with the trend that the magnitude of change in the light output decreases as the proton recoil energy increases.

For a measurement in which the proton recoils travel through the crystal at a range of angles, this directional dependence introduces additional variability into the light output and pulse shapes produced by neutron interactions in anthracene. A metric \sanis that quantifies the anisotropy in terms of this resolution effect is related to the observed standard deviation \sobs of measured \LOave and \TTTave values at a given energy. The contribution from statistical variance is subtracted in quadrature, e.g.\
\begin{equation*}
\sanis=\sqrt{\sobs^{2} - \sstat^{2}},
\end{equation*}
where $\sstat^{2}$ is the average statistical variance from the set of measurements at different recoil directions. The relevant quantities for each group of datasets are given in \tab{Table:AnthNeutronVariances}, normalized to the average measured value $\mu$.

\begin{table}[ht]
\centering
\caption{Variability in Pulse Shape and Light Output in Proton Recoil Events in Anthracene.}
\label{Table:AnthNeutronVariances}
\begin{tabular}{c|c|r|r}
                            & \ERecoil      & 14 MeV & 2.5 MeV \\
    \hline
	\multirow{3}{*}{\LOave} & \sobs$/\mu$   & 3.7\%   & 8.6\% \\    
	                        & \sstat$/\mu$  & 0.4\%   & 2.2\% \\      
	                        & \sanis$/\mu$  & 3.7\%   & 8.3\% \\      
	\hline
	\multirow{3}{*}{\TTTave}& \sobs$/\mu$   & 16.8\%  & 7.2\% \\    
	                        & \sstat$/\mu$  & 0.3\%   & 0.4\% \\      
	                        & \sanis$/\mu$  & 16.8\%  & 7.2\%
\end{tabular}
\end{table}

Strictly speaking, \sanis includes both the anisotropy effect and other sources of systematic error. These systematic errors include but are not limited to the width of the proton recoil direction selection window due to the physical size of the detector and the range of energies selected at the endpoint, the possibility of several low energy interactions summing to a full energy event, the polycrystallinity that exists in the sample, and the approximation of the fit function to represent the distribution. Given efforts to limit these systematic biases and the qualitative features in the angular distribution of \LOave and \TTTave, it appears that \sanis for both \LOave and \TTTave are dominated by the anisotropy effect in proton recoil interactions. 

Other explanations for the observed directional dependence have been considered, such as magnetic field effects on the PMT. However, any such external effect must be minimal compared to the internal effect in the crystal because this anisotropy is not observed for gamma-ray interactions, as will be demonstrated in \secref{Sec:Gamma-Ray Measurements}.

\subsection{Discussion}\label{Subsec:NeutronDiscussion}
According to measurements made by previous groups, the proton recoil direction that produces maximum light output in anthracene is along the $c'$-axis, defined as the direction perpendicular to the $ab$-plane of the crystal, and the direction that produces minimum light output is along the $b$-axis~\cite{Brooks197469,Tsukada1962286}. This allows for the crystal axes of the anthracene sample measured in this paper to be inferred. According to Figs.~\ref{Fig:2015_02_12_Anth_DT} and~\ref{Fig:2015_02_23_Anth_DD}, the $b$-axis is at approximately $(\theta,\phi)=(40^\circ,330^\circ)$ and the $c'$-axis is at approximately $(\theta,\phi)=(60^\circ,200^\circ)$. These directions make an angle of approximately $90^\circ$, as the $b$ and $c'$-axes should. This puts the $a$ axis at approximately 
$(\theta,\phi)=(65^\circ,95^\circ)$, which is roughly the position of the saddle point observed in the \TTTave distribution. These measurements also confirm Tsukada's findings that the saddle point in the light output distribution occurs in the $ac$ plane about $30^\circ$ from the $a-$axis~\cite{Tsukada1962286}.

Brooks has hypothesized that the directional dependence in heavy charged particle events is a result of preferred directions of excitation transport in the crystal~\cite{Brooks197469}. Excitation transport changes the excitation density over time in the material. In the same way that the light output and pulse shape differ for neutron and gamma-ray events due to the gross differences in the excitation densities that they produce, as described in \secref{Sec:LightEmission}, small changes in the excitation density due to directionally dependent transport within the material may affect the light output and pulse shape at the levels observed in these measurements.

The evolution of the excitation density depends on a number of things, one of which is the initial distribution of excitations with respect to the preferred directions of transport. If the excitations are deposited along a direction with rapid transport, they are more likely to move within the initial path, maintaining the excitation density and the probability that excitations will meet and interact throughout their lifetime. If the excitations are deposited perpendicular to the direction with rapid transport, the excitations more easily transport away from the path during their lifetime, decreasing the excitation density and reducing the probability to interact.

The overall light output \LOave depends primarily on the rate of singlet quenching, while the pulse shape parameter \TTTave is affected by both singlet quenching and triplet annihilation. As evident in \fig{Fig:2015_02_12_Anth_DT} and \fig{Fig:2015_02_23_Anth_DD}, the proton recoil directions that produce higher light output correspond to regions that produce lower pulse shape values, and vice versa. This may lead one to conclude that a change in \LOave is accompanied by an opposite change in \TTTave, but it is apparent from the different positions of the saddle points that the \LOave and \TTTave values are not directly coupled. This correlation, but not fixed relationship, between the \LOave and \TTTave distributions indicates that the physical mechanisms responsible for the directional dependence may affect singlet and triplet state evolution in related but different ways. The singlet and triplet excitations undergo transport and interactions via different mechanisms, so they may or may not have the same preferred directions of transport or the same relative change in transport likelihood in different directions. The same may be true of other kinetic processes. This means that singlet and triplet excitations will experience different changes in their densities throughout their lifetimes. 

It is possible that the magnitude of change in the excitation density due to preferred directions of transport is only significant among a high excitation density like that produced by neutron events. This hypothesis will be tested through measurements of gamma-ray and cosmic muon interactions in the following sections.

\section{Gamma-Ray Measurements}\label{Sec:Gamma-Ray Measurements}

\subsection{Purpose of Gamma-Ray Measurements} 
So far, no directional dependence has been documented in gamma-ray interactions in organic crystal scintillator detectors. Brooks' characterization of the heavy charged particle scintillation anisotropy noted that electron recoil events produced by gamma-ray interactions are not subject to a directional dependence~\cite{Brooks197469}, however, this was not quantified and no measurement details were provided.


According to the working hypothesis on the mechanisms that cause the directional dependence, the excitation density produced by gamma-ray events may be so low that directionally dependent changes are not significant compared to the overall density. However, it is also possible that the effect is just much smaller for gamma-ray events than for neutron events, and a smaller anisotropy may be observable with a more sensitive measurement. In order to investigate this, electron recoils produced by gamma-ray interactions have been measured at different directions in anthracene.

\subsection{Gamma-Ray Interactions in Anthracene} 
Gamma rays of energies typical of radioactive sources interact primarily via Compton scattering in anthracene. For an incident gamma ray of energy $E_{\gamma}$, the highest energy electron recoil that can be produced occurs in the head-on collision in which the photon scatters at angle $\theta=\pi$ from its initial direction and the recoil electron departs the interaction in the forward direction. Although the initial direction of the electron is fixed for such a Compton edge event, the electron undergoes wide angle scattering as it travels through the crystal, so it does not travel in a straight path as it deposits its energy.

\subsection{Experimental Setup}
Measurements were made with a Cs-137 source that was placed 38" from the detector, controlling the incident gamma-ray direction within 3$^\circ$. The 662 keV gamma ray from Cs-137 produces a Compton edge electron recoil with energy $E_{e^-}=478$ keV. In order to control the energy and initial direction of electron recoils produced in anthracene, Compton edge electron events were selected. The same rotational stage was used for the gamma-ray measurements as was described in \secref{Sec:NeutronExperimentalSetup} for the neutron measurements. This provided the capability to change the initial direction of the electron recoil in the crystal axes by rotating the anthracene crystal with respect to the incident gamma-ray direction. Measurements were made at the same 72 recoil directions in the anthracene crystal as were measured in the DT neutron measurements. 

\subsection{Data Analysis and Results} 
For each event, the \LO value was calculated in summed digitizer channel units using \eqnref{Eqn:LightOutput} with a dimensionless $C=1$. In this analysis, no conversion to keVee was made. In order to build a fit function, MCNP5 was used to produce a simulated detector response. The energy spectrum produced by 662 keV gamma rays incident on anthracene in MCNP5 was smeared with a Gaussian detector response function with resolution $\sigma$, which was adjusted until the smeared spectrum matched the shape from measured light output spectra, as shown in \fig{Fig:Cs137_Fit_Function_Demonstration}. This simulated spectrum showed that the amplitude of the energy spectrum at 478 keV was equal to $N_{478}=0.74*N_{CE}$, where $N_{CE}$ is the amplitude of the energy spectrum at the peak corresponding to the Compton edge. 

For each measurement, the following three steps were taken to calculate \LOave and \TTTave, the expected \LO and \TTT values for a 478 keV electron recoil at the angle of interest.

\begin{figure}
	\centering
	\includegraphics[trim={0cm 0cm 0cm 0cm},clip,scale=.8]{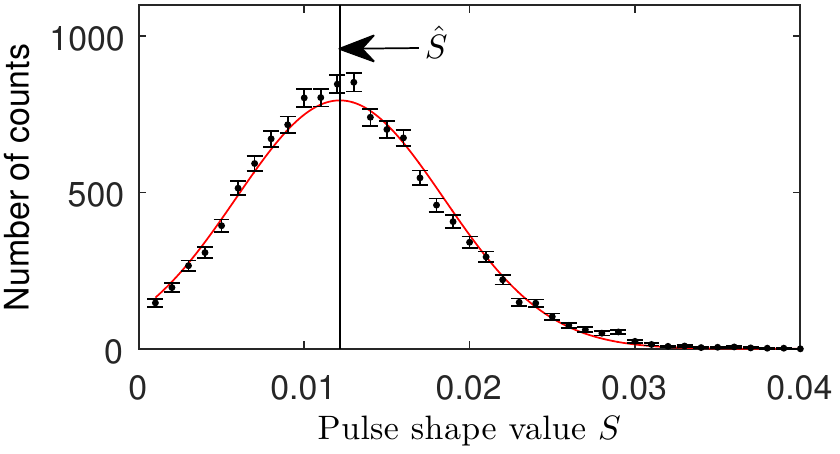}
	\caption{Distribution of \TTT values for events with light output in the range \LOaverange produced by Cs-137 gamma rays incident on anthracene at $(\theta,\phi)=(50^\circ,178^\circ)$. Black points are experimental data with statistical error bars. The red line is the applied Gaussian fit function.
	\label{Fig:Cs137_TTT_Distribution}
	}
\end{figure}

\begin{enumerate}
\item Light output spectrum fit: A light output spectrum of all events in a single measurement is shown in \fig{Fig:Cs137_Fit_Function_Demonstration}. The light output corresponding to $N_{478}$ was recorded as \LOave.

\begin{figure}
	\centering
	\includegraphics[trim={0cm 0cm 0cm 0cm},clip,scale=.8]{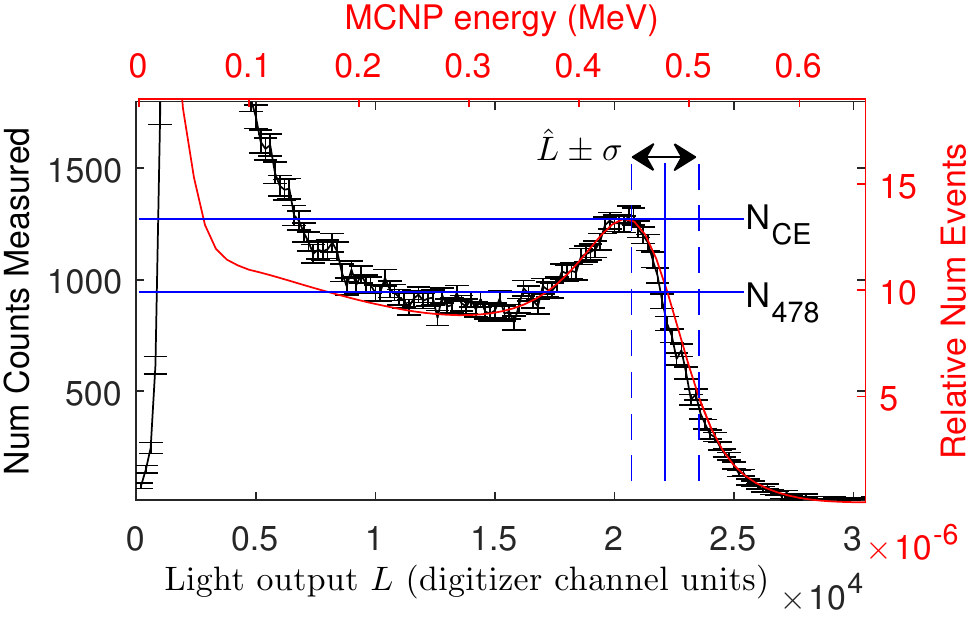}
	\caption{Light output spectrum for Cs-137 gamma-ray events incident on anthracene. The red curve and axes correspond to an MCNP5 simulation. The black data points and axes correspond to a measurement. $N_{478}$ and $N_{CE}$ as found in the fit function are indicated, which produce a final \LOave value for the light output in summed digitizer channel units of a 478 keV electron recoil event.
	\label{Fig:Cs137_Fit_Function_Demonstration}
	}
\end{figure}

\item Compton edge event selection: Events within the range \LOaverange were selected as Compton edge electron recoils. This widened the range of electron recoil directions to events within half-angle $\rho=18.3^\circ$ around the forward direction.

\item Pulse shape distribution fit: A distribution of the \TTT value for Compton edge electron recoil events was produced. A Gaussian fit was applied to this distribution to estimate the expected pulse shape value \TTTave as shown in \fig{Fig:Cs137_TTT_Distribution}.
\end{enumerate}

In order to estimate the statistical error in the calculation of \LOave in each measurement, a bootstrapping method was applied. 100 light output spectra were generated based on Poisson fluctuations about the light output spectrum from each measurement, simulating a resampling of the data. The \LOave value was calculated for each bootstrap-generated spectrum, and the standard deviation in the 100 \LOave values served as an estimate for the statistical error in the measurement of \LOave.

\begin{figure}
	\centering
	\includegraphics[width=2.5in]{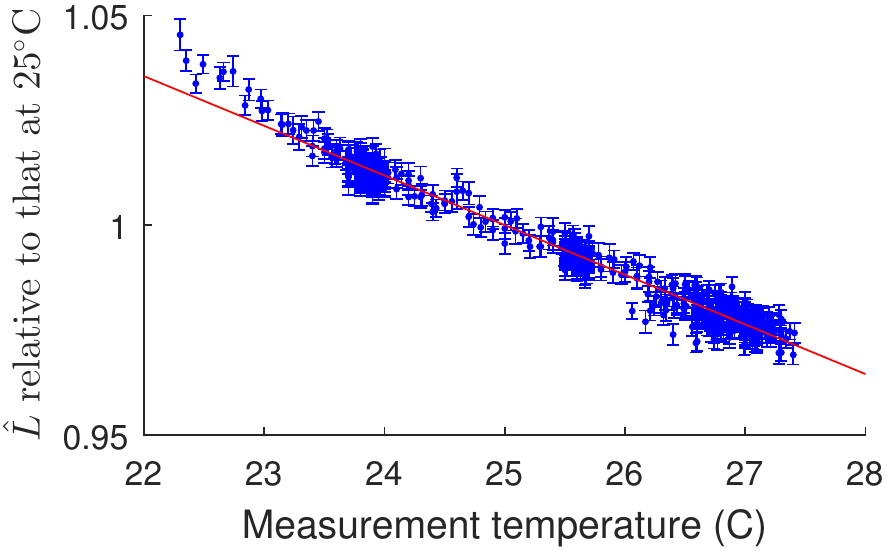}
	\caption{\LOave produced by a Cs-137 source at a fixed position at temperatures 22-28$^{\circ}$C relative to that at 25$^\circ$C.
	\label{Fig:2015_08_18_Lave_vs_Temp_Rel}	
	}
\end{figure}

\begin{figure}
	\centering
	\includegraphics[width=2.5in]{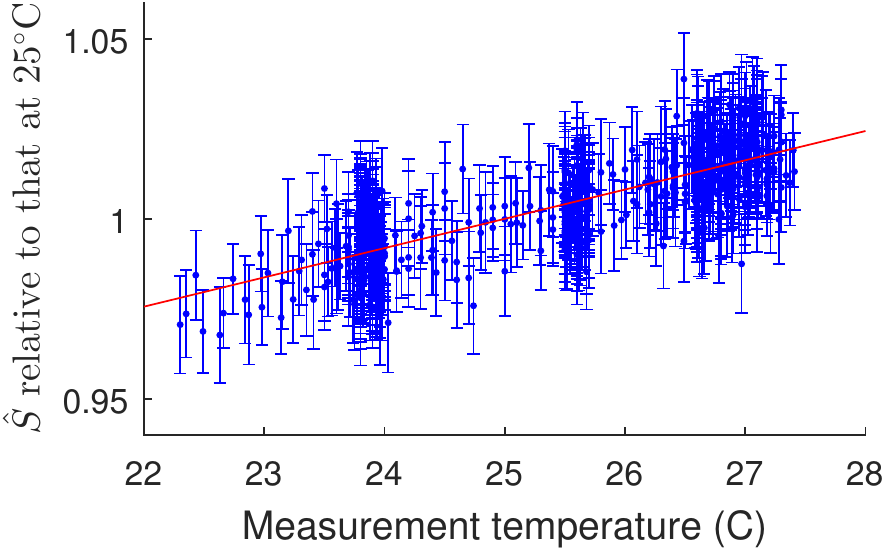}
	\caption{\TTTave produced by a Cs-137 source at a fixed position at temperatures 22-28$^{\circ}$C relative to that at 25$^\circ$C.
	\label{Fig:2015_08_18_Save_vs_Temp_Rel}
	}
\end{figure}

The light output and pulse shapes produced in anthracene have a temperature dependence that proved to be the largest source of systematic bias in the gamma-ray measurements. Therefore, the dependence was characterized and a correction was applied. A separate set of 10-min long measurements of a Cs137 source at a fixed angle with respect to the detector were taken over six days as the temperature in the lab varied with the weather. The \LOave and \TTTave values were calculated for each measurement and plotted vs.\ temperature as shown in \fig{Fig:2015_08_18_Lave_vs_Temp_Rel} and \fig{Fig:2015_08_18_Save_vs_Temp_Rel}. Linear fits to these data were used to correct the \LOave and \TTTave values in the directional measurements to 25$^\circ$C.

\begin{figure}[!t]
\centering
\subfloat[Light output \LOave.]{\includegraphics[trim={0cm .7cm 0cm 0cm},clip,width=2.5in]{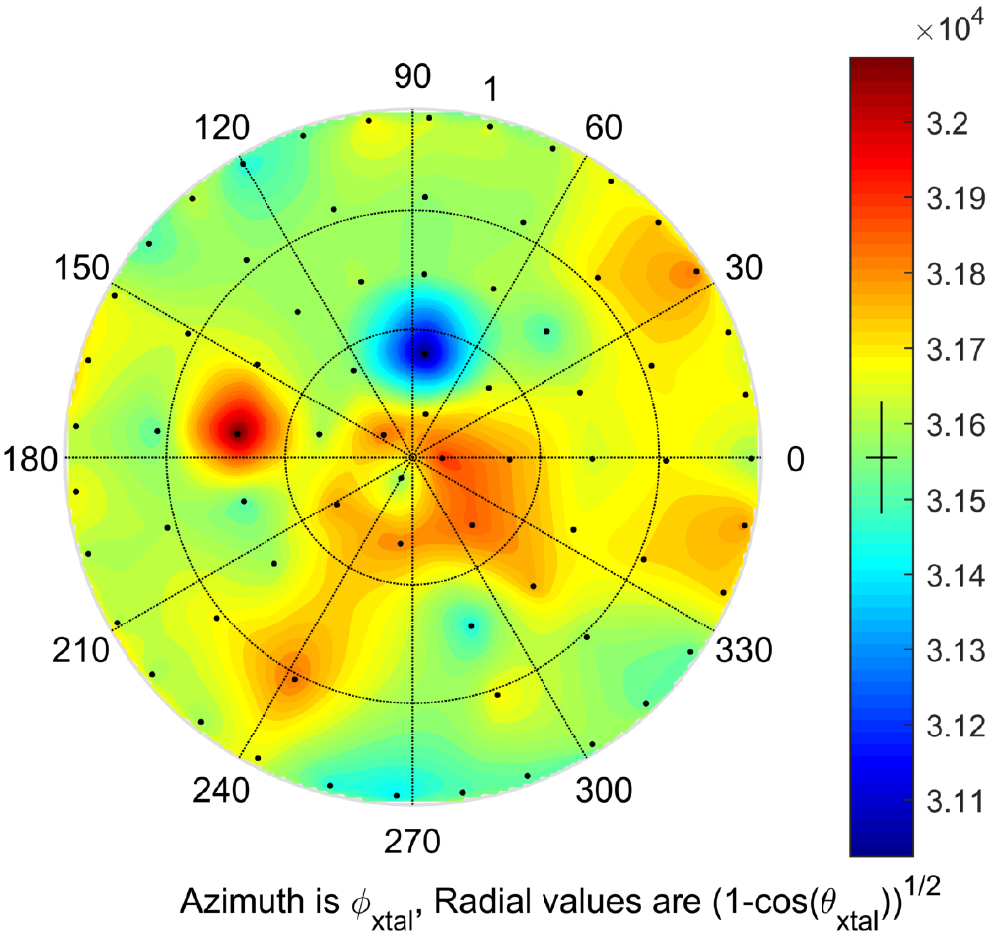}%
\label{Fig:Cs137_Full_Space_LO_PolarPlot}}
\hfil
\subfloat[Pulse shape value \TTTave.]{\includegraphics[trim={0cm .7cm 0cm 0cm},clip,width=2.5in]{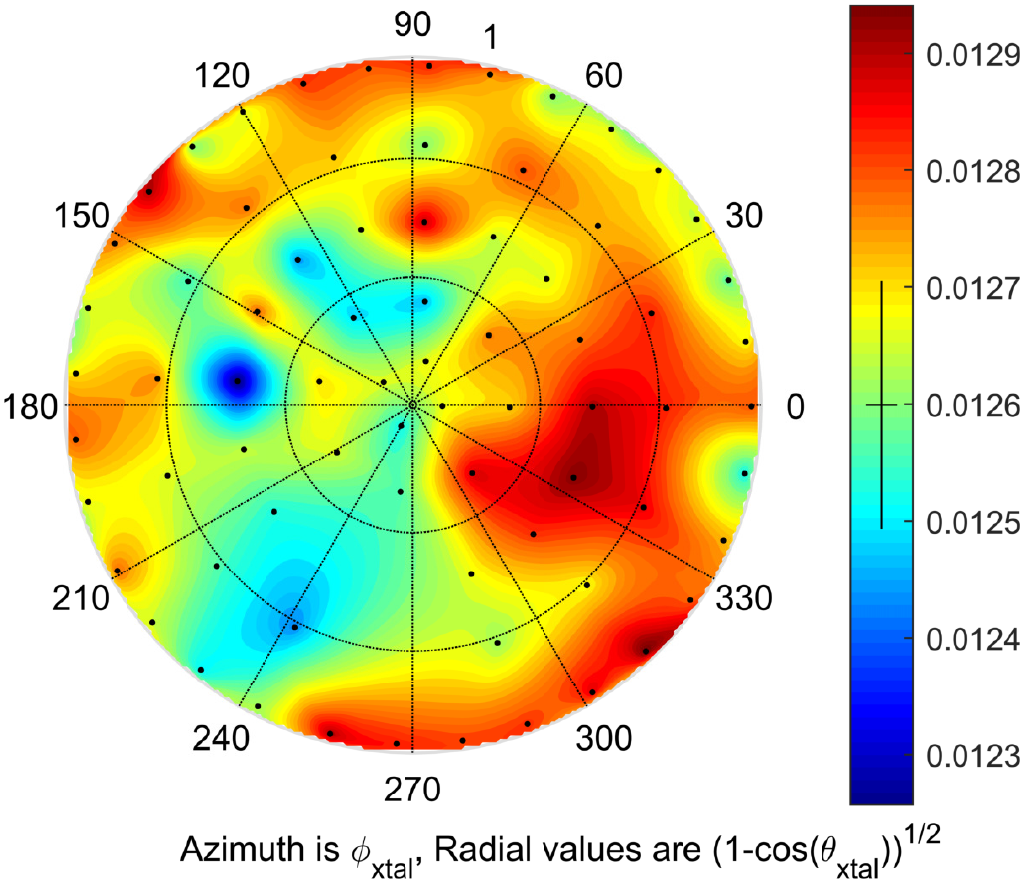}%
\label{Fig:Cs137_Full_Space_TTT_PolarPlot}}
\caption{Response of anthracene crystal at various recoil directions for 478 keV electron recoils. Black points indicate measurements and the colors represent a smooth interpolation between measurements. Length of vertical black bar on colorbar indicates average 2-$\sigma$ statistical error.}
\label{Fig:Cs137_PolarPlots}
\end{figure}

\fig{Fig:Cs137_PolarPlots} shows the \LOave and \TTTave values measured for 478 keV electron recoil events in anthracene at different electron recoil directions. The length of the black line on the colorbar is the average statistical uncertainty in the measurement. Three measurements were omitted in which the motor system was between the source and the detector, causing considerably more environmental scattering and producing outlying values for \LOave and \TTTave.

Although there are variations in these measurements greater than the statistical uncertainties, there is not a distinct pattern of high and low regions in angle space as there are in the proton recoil measurements shown in \secref{Sec:NeutronDataAnalysis}. Since the variation has no order, it appears that there is no measurable directional dependence, but rather there are other sources of variation.

Following the method explained in \secref{Sec:NeutronDataAnalysis}, \tab{Table:AnthGammaVariances} shows the relative standard deviations due to statistical uncertainty and other effects. Compared to the neutron measurements in which it was concluded that the anisotropy was the dominant effect in the variability of \LOave and \TTTave, the qualitative features on the angular distributions of \LOave and \TTTave indicate that the variability in the gamma-ray measurements is dominated by sources of bias and not an anisotropy. For that reason, this variability will be named \sother. Several sources of bias exist in these measurements. First, as the rotational stage changes its position, the position of the detector relative to the source and environment change, so the scattering environment differs. This may be significant for a gamma-ray source among the high-Z materials in the laboratory equipment. Second, the temperature correction is not a perfect method because the temperature sensor is outside of the detector and provides a measurement of the room temperature rather than the temperature in the crystal. Third, the light output window in the gamma-ray measurements produces a wider selection of recoil angles than in the neutron measurements.  

\begin{table}[ht]
\centering
\caption{Variability in Pulse Shape and Light Output in Electron Recoil Events in Anthracene.}
\label{Table:AnthGammaVariances}
\begin{tabular}{c|c|r}
                            & \ERecoil             & 0.478 MeV\\
    \hline
	\multirow{3}{*}{\LOave} & \sobs/$\mu$   & 0.5\%   \\    
	                        & \sstat/$\mu$  & 0.3\%   \\      
	                        & \sother/$\mu$ & 0.4\%   \\      
	\hline
	\multirow{3}{*}{\TTTave}& \sobs/$\mu$   & 1.1\%   \\    
	                        & \sstat/$\mu$  & 0.8\%   \\      
	                        & \sother/$\mu$ & 0.7\%  
\end{tabular}
\end{table}

However, if a conservative assumption is made that the anisotropy is the dominating effect, an upper limit on the magnitude of the anisotropy from 478 keV electron recoils may be approximated as \sother. Even under this assumption, the anisotropy effect in electron recoil events is still less than one tenth of that for the proton recoil interactions (cf.\ \tab{Table:AnthNeutronMagnitudeChange}).

\subsection{Discussion} 
This analysis provides quantitative measurements that support the claim made by previous groups that no directional dependence has been observed in electron recoil events in anthracene~\cite{Brooks197469}. Two reasons have been hypothesized as being responsible for the lack of directional dependence in electron recoil events. First, Brooks attributed this lack of anisotropy to the non-straight path that electrons follow as they slow down due to the large-angle scattering that they undergo~\cite{Brooks197469}. Electrons may not actually populate a directionally dependent excitation distribution, and it is possible that an electron recoil event, were it to travel in a straight path, could be subject to a directional dependence, but the non-straight path traveled by the electron washes out the effect. Second, electrons deposit their energy with a much lower \dEdx than heavy charged particles, producing a lower excitation density. Changes in the excitation density due to directional transport may be on too small a scale compared to the overall density to affect the relative rates of kinetic processes for the electron recoil. The cosmic muon measurements presented in \secref{Sec:Cosmic Muon Measurements} provide a test of these hypotheses.

\section{Cosmic Muon Measurements}\label{Sec:Cosmic Muon Measurements}

\subsection{Purpose}\label{Sec:MuonPurpose} 
As demonstrated above, a directional dependence in anthracene has been observed in heavy charged particle interactions but not in electron recoil events. In order to test whether the lack of directional dependence from the electron recoil is due to its low \dEdx or due to its non-straight path, cosmic muon events were measured. Muons are elementary particles similar to electrons but with mass of 105.7 MeV/c$^2$. Like electrons, muons interact with a much lower \dEdx than heavy charged particles. However, due to their large mass compared to the electrons in the medium in which they are interacting, muons are not subject to large-angle scattering and thus travel in a quasi-straight path.

Since muons travel with lower \dEdx like the electron recoil, but in a quasi-straight path like the proton recoil, the presence or lack of directional dependence in muon interactions will provide information on whether a directional dependence requires that a particle interact with high \dEdx or in a straight path. The goal of these measurements is to measure the anisotropy present in muon events in anthracene, or if no anisotropy is observed, set an upper bound on its magnitude.

\subsection{Muon Interactions in Anthracene} 
Cosmic muons reach sea level with approximately 4 GeV energy. They interact as minimally ionizing particles, depositing a minimal amount of energy per distance across the length of the material in which it interacts. In anthracene, muons deposit approximately 2.4 MeV/cm\cite{lbnlmuon}. This is very close to the \dEdx deposited by a 478 keV electron of 2.5 MeV/cm, and much less than the 166.7 MeV/cm deposited by a 2.5 MeV proton recoil\cite{nist}. Because a muon's mass is so much greater than an electron's mass, a muon experiences minimal changes in its direction as it interacts with electrons in the medium, producing a quasi-straight path.

Since the energy deposited is proportional to the path length that the muon travels in the detector, the deposited energy spectrum will be equal to the path length distribution multiplied by a constant factor. If there were a light output anisotropy in muon interactions in anthracene, it is expected that muons traveling at different angles would produce a different light output per energy deposited. Thus, the light output spectra produced by muons traveling at different directions would be equal to the path length distribution multiplied by different constant factors, and any features in those spectra would be shifted to different light output values.

\subsection{Experimental Setup} 
\begin{figure}[!t]
\centering
\subfloat[Cartoon of experimental setup (not to scale)]{\includegraphics[width=1.5in]{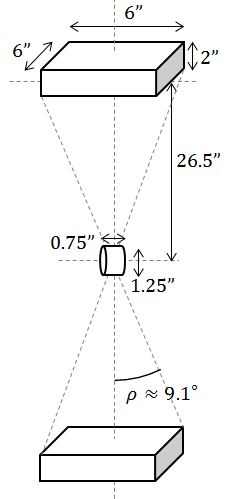}%
\label{Fig:MuonExperimentalSetupCartoon}}
\hfil
\subfloat[Photo of experimental setup]{\includegraphics[width=1.5in]{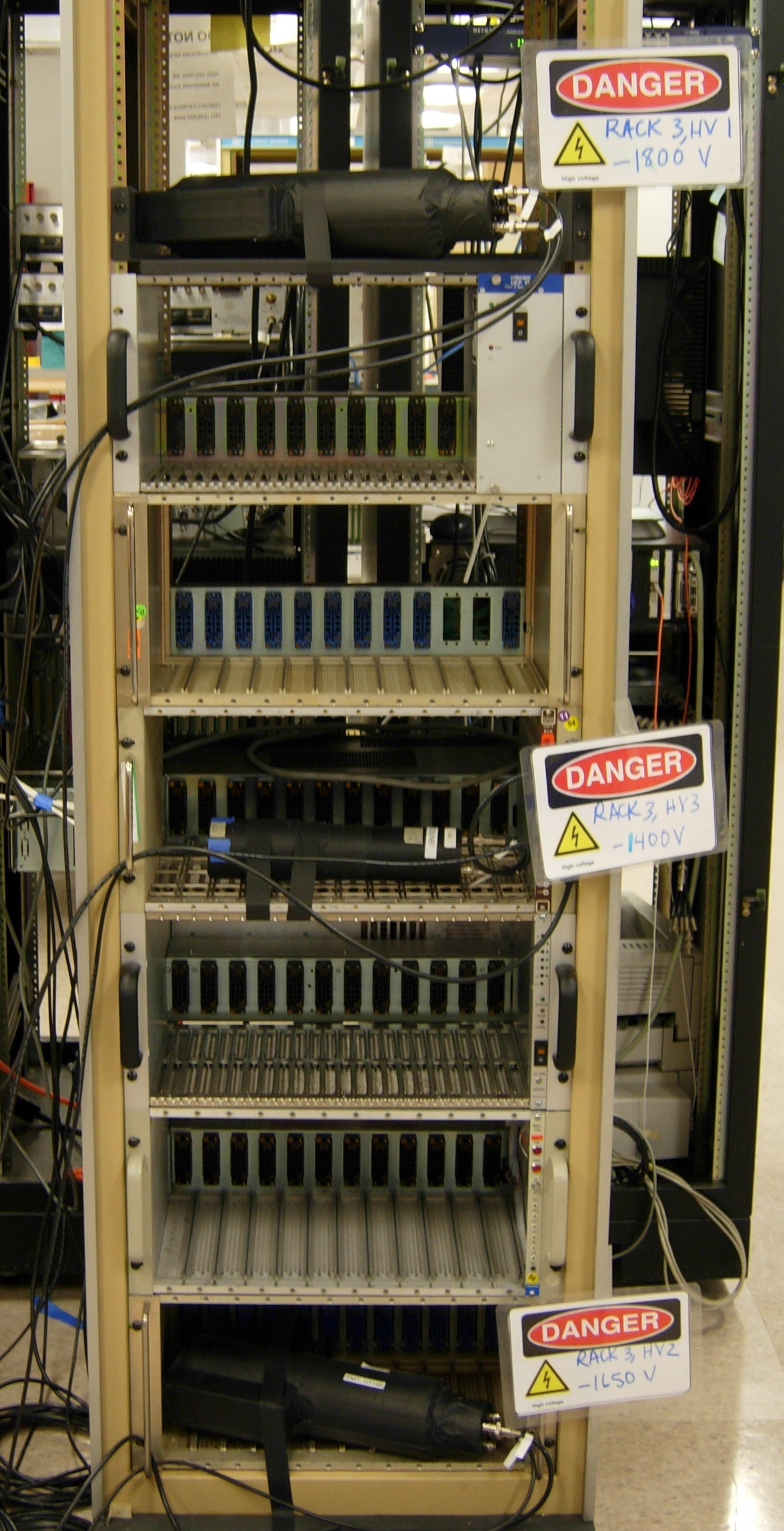}%
\label{Fig:MuonExperimentalSetupPhoto}}
\caption{Experimental setup used in cosmic muon measurements.}
\label{Fig:MuonExperimentalSetup}
\end{figure}

\fig{Fig:MuonExperimentalSetup} shows the detectors used in this measurement. The rectangular blocks are EJ200 plastic scintillator detectors, and the cylinder is the anthracene crystal. Only events that exceeded the trigger threshold in all three detectors were used in order to select muons that traveled within a set angle $\rho$ from the vertical direction through the anthracene. As the distance between the anthracene and the plastic blocks is increased, the angle $\rho$ decreases to select muons traveling within a narrower range of angles in the anthracene, and the count rate decreases. The distance was chosen so that $\rho$ was comparable to the range of proton recoil directions accepted in the neutron measurements. Each plastic block was placed 26.5" away from the anthracene detector, limiting the muon directions to within the half-angle $\rho=9.1^\circ$ of the vertical direction.

In order to investigate muon interactions at different directions within the anthracene crystal axes, measurements were taken with the crystal at different orientations with respect to a vertical muon trajectory. Due to the requirement that the geometry of the system be identical in the measurements in order to preserve the path length distribution, only angles at which the height axis of the crystal was perpendicular to the vertical muon path, as shown in \fig{Fig:MuonExperimentalSetupCartoon}, were candidates for the muon measurements. Two such directions were chosen and will be referred to as directions 1 and 2. Direction 1 was selected at $(\theta,\phi)=(90^\circ,68.7^\circ)$, and direction 2 was at $(\theta,\phi)=(90^\circ,158.7^\circ)$. Each measurement was taken for 20 days, producing approximately 2300 muon events in each measurement. 

An assumption has been made that any anisotropy that would exist in the light output and pulse shape produced by muon interactions would follow the same crystal axes as the anisotropy from proton recoil events. Although these two interaction directions are not those of greatest difference in light output and pulse shape from the neutron measurements, there was still a significant difference in the light output and pulse shape at these two angles from 14 MeV and 2.5 MeV proton recoils, as listed in \tab{Table:MuonResults}.

\subsection{Data Analysis and Results}
\begin{figure}
	\centering
	\includegraphics[trim={0cm 0cm 0cm 0cm},clip,width=2.2in]{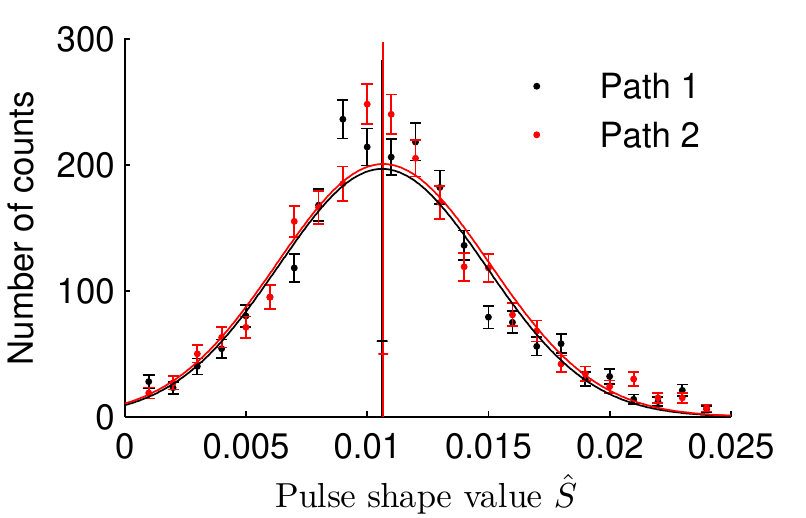}
	\caption{Distribution of \TTT values for events with light output in the range \LOaverange produced by cosmic muons in anthracene. The overlying curves are the Gaussian fits applied, and the vertical lines are the location of the expected pulse shape values \TTTave.
	\label{Fig:MuonTTTDistributions}
	}
\end{figure}

In order to evaluate whether there is a significant difference in the light output and pulse shapes produced by muon events at directions 1 and 2, the expected light output \LOave and pulse shape value \TTTave for events at the peak feature in the light output spectrum were calculated by following these three steps:

\begin{enumerate}
\item Light output spectrum fit: A light output spectrum of all events was produced, as shown for the measurement at direction 1 in \fig{Fig:MuonLOSpectrum1}. A Gaussian fit with centroid \LOave and standard deviation $\sigma$ was applied to the peak feature, as shown for both directions in \fig{Fig:MuonLOPeakFit}.

\begin{figure}[!t]
\centering
\subfloat[Light output spectrum for muon events through anthracene at angle 1.]{\includegraphics[trim={0 0 0 0.5cm},clip,width=1.5in]{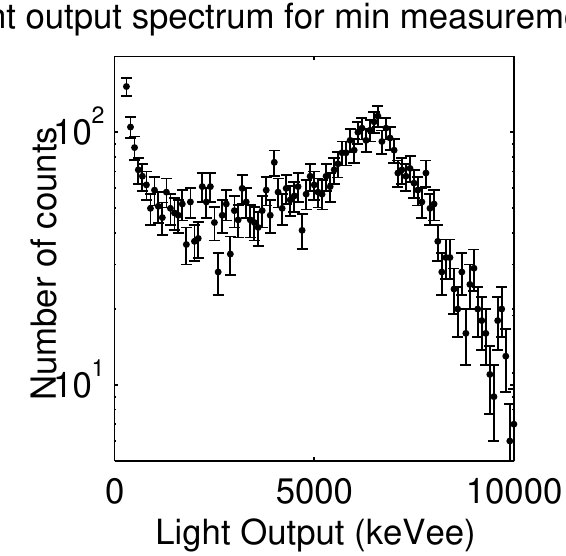}%
\label{Fig:MuonLOSpectrum1}}
\hfil
\subfloat[Comparison of peak in light output spectra for muons traveling at directions 1 and 2. The vertical line indicates \LOave values as calculated by Gaussian fits.]{\includegraphics[trim={0 0 0 0.4cm},clip,width=1.5in]{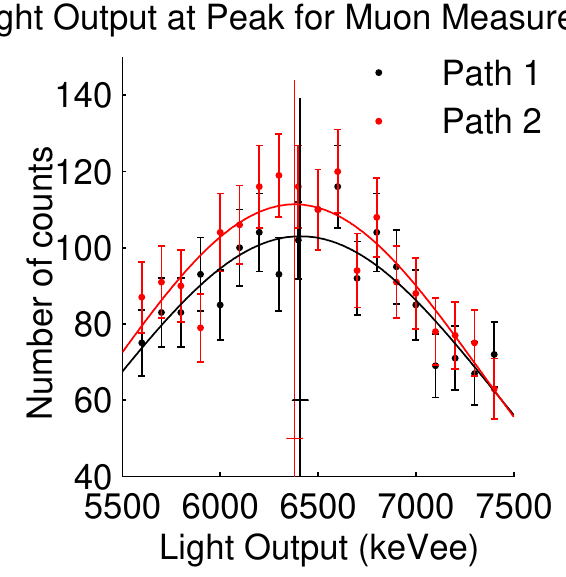}%
\label{Fig:MuonLOPeakFit}}
\caption{Analysis of light output spectra in muon measurements.}
\label{Fig:MuonLOAnalysis}
\end{figure}
\item Peak event selection: Events in the peak feature are identified by selecting events with light output in the range \LOaverange. 

\item Pulse shape distribution fit: A distribution of the \TTT value for peak events is produced. A Gaussian fit is applied to this distribution to estimate the expected pulse shape value \TTTave produced by a muon, as shown in \fig{Fig:MuonTTTDistributions}.
\end{enumerate}

The magnitudes of change in the \LOave and \TTTave between measurements were calculated as the ratio of the maximum to minimum of each measurement. The magnitude of change in the \LOave value was $1.005\pm0.009$, and the magnitude of change in the \TTTave value was $1.004\pm0.026$. Neither showed a statistically significant change for muons between paths 1 and 2. 

It is still possible that there is a very small anisotropy present that is not measurable by this system. These results can serve to set an upper boundary on the magnitude of anisotropy in anthracene at these muon paths. To 1-$\sigma$, these results are inconsistent with a magnitude of change in light output greater than 1.013 and a magnitude of change in the pulse shape value greater than 1.030 for muons traveling at directions 1 and 2.

\subsection{Discussion}
\tab{Table:MuonResults} compares the magnitude of change measured in the light output and pulse shape value for protons, electrons, and muons traveling at directions 1 and 2 in the anthracene crystal. The table shows two major differences in the interactions of these particles. First, protons and muons travel in a straight path as they deposit their energy, while the electron does not. Second, the electron and muon produce comparable \dEdx, while the proton recoil produces much higher \dEdx in the material. Of these particles, an anisotropy was only observed in proton recoil interactions.

\begin{table*}
\centering
\caption{Summary of Measurements Made on Anthracene Sample for Interactions at Directions 1 and 2.}
\label{Table:MuonResults}
\small
\begin{tabular}{lllll}
Source Particle      & Neutron          & Neutron          & Gamma ray       & Muon             \\
Recoil Particle      & Proton           & Proton           & Electron        & --               \\
Path                 & Straight         & Straight         & Non-straight    & Straight         \\
Energy (MeV)         & 14               & 2.5              & 0.478           & 4000             \\
dE/dx (MeV/cm)       & 42.9             & 166.7            & 2.5       & 2.4\\
\LOave mag. change   & 1.060$\pm$0.005  & 1.108$\pm$0.012  & 1.005$\pm$0.004 & 1.005$\pm$0.009  \\
\TTTave mag. change  & 1.062$\pm$0.002  & 1.022$\pm$0.007  & 1.006$\pm$0.017 & 1.004$\pm$0.026  \\
Anisotropy Observed  & Yes              & Yes              & No              & No      
\end{tabular}
\end{table*}

The lack of anisotropy observed in muon interactions provides new insights on the mechanism that produces the directional dependence in heavy charged particle interactions. Since the muon, which travels in a straight path, does not experience a directional dependence, it can be concluded that the meandering recoil electron is not solely responsible for the lack of directional dependence in gamma-ray interactions. This lends support to the theory that a high \dEdx is necessary for producing a directional dependence. 

This result agrees with the hypothesis presented in \secref{Subsec:NeutronDiscussion} that says the anisotropy is partly due to preferred directions of excitation transport. This transport changes the excitation density over time. Depending on the initial distribution of excitations compared to the directions of rapid transport, the excitations may move towards one another or away from one another, changing the rates of interactive processes such as triplet-triplet annihilation and singlet quenching and, in turn, affecting the amount of light produced and the time distribution. For heavy charged particle that interact with high \dEdx, the change in the excitation density is significant compared to the overall density. For gamma-ray and muon interactions, the overall excitation density is low enough due to their low \dEdx that these changes are not significant enough to change the scintillation output on an observable level.

\section{Conclusion}
The anisotropic scintillation response of crystalline anthracene to heavy charged particles has been investigated through a series of measurements using incident neutrons, gamma rays, and muons. The directional dependence of the scintillation amplitude and pulse shape for proton recoils at 14 MeV and 2.5 MeV is consistent with previous measurements at similar energies. These measurements are used to evaluate the contribution of the anisotropy to the energy resolution of anthracene for typical neutron detection scenarios in which the proton recoil direction is not known event by event. This contribution is 3.7\% for 14 MeV proton energy deposited and 8.3\% for 2.5 MeV proton energy deposited.

In identical measurements using incident 662~keV gamma rays, an an\-i\-so\-trop\-ic response was not observed. Variations among the datasets at different electron recoil angles do not follow a clear directional pattern, and are consistent with unrelated systematic variability. But under the conservative assumption that observed variability is due to an anisotropy, we limit the relative magnitude of the directional variability to less than a tenth of that observed in the neutron measurements. This is the first quantitative estimate of or limit on the anisotropic response of a crystal organic scintillator to electron recoils.

Finally, in order to aid in distinguishing among different hypotheses for the physical origin of the anisotropy, cosmic muons were measured at two directions through the same anthracene crystal. These measurements are statistically limited and only one pair of directions was measured, but the absence of a significant difference in light output or pulse shape between the two directions indicates that the anisotropy is weak or not present for low-\dEdx particles, even when the particle track is straight.

Beyond preferred directions of excitation transport, it is unclear what physical mechanisms contribute to the anisotropic scintillation response in neutron interactions, and what physical or chemical properties dictate its magnitude and behavior in a given material. This work raises the question whether the effect may be corrected for in order to improve energy resolution, and whether the effect may be exploited to use these materials as compact directional neutron detectors. Both of these applications would benefit from a deeper understanding of the physical mechanism that is responsible for the anisotropy.  

%
\section*{Acknowledgements}
The authors wish to thank John Steele for his assistance in building the motor driven rotational stage, and Andrew Glenn of Lawrence Livermore National Laboratory for his suggestion to measure cosmic muon events.

This material is based upon work supported by the National Science Foundation Graduate Research Fellowship Program under Grant No. DGE 1106400. This material is based upon work supported by the Department of Energy National Nuclear Security Administration under Award Number. DE-NA0000979 through the Nuclear Science and Security Consortium. Sandia National Laboratories is a multi-program laboratory managed and operated by Sandia Corporation, a wholly owned subsidiary of Lockheed Martin Corporation, for the U.S. Department of Energy's National Nuclear Security Administration under contract DE-AC04-94AL85000. 

%


\begin{thebibliography}{1}
\expandafter\ifx\csname url\endcsname\relax
  \def\url#1{\texttt{#1}}\fi
\expandafter\ifx\csname urlprefix\endcsname\relax\def\urlprefix{URL }\fi
\expandafter\ifx\csname href\endcsname\relax
  \def\href#1#2{#2} \def\path#1{#1}\fi

\bibitem{Knoll}
G.~Knoll, Radiation Detection and Measurement, Wiley, New York, 2000.

\bibitem{Carman201356}
L.~Carman, N.~Zaitseva, H.~P. Martinez, B.~Rupert, I.~Pawelczak, A.~Glenn,
  H.~Mulcahy, R.~Leif, K.~Lewis, S.~Payne, The effect of material purity on the
  optical and scintillation properties of solution-grown trans-stilbene
  crystals, Journal of Crystal Growth 368 (2013) 56 -- 61.
\newblock \href
  {http://dx.doi.org/http://dx.doi.org/10.1016/j.jcrysgro.2013.01.019}
  {\path{doi:http://dx.doi.org/10.1016/j.jcrysgro.2013.01.019}}.

\bibitem{Brooks197469}
F.~Brooks, D.~Jones, Directional an\-i\-sot\-ro\-py in organic scintillation
  crystals, Nuclear Instruments and Methods 121~(1) (1974) 69 -- 76.
\newblock \href
  {http://dx.doi.org/http://dx.doi.org/10.1016/0029-554X(74)90141-4}
  {\path{doi:http://dx.doi.org/10.1016/0029-554X(74)90141-4}}.

\bibitem{birks1964theory}
J.~Birks, The Theory and Practice of Scintillation Counting, International
  series of monographs on electronics and instrumentation, Pergamon Press,
  Oxford, 1964.

\bibitem{Azumi1963}
T.~Azumi, S.~P. McGlynn, Delayed fluorescence of solid solutions of polyacenes.
  ii. kinetic considerations, The Journal of Chemical Physics 39~(5) (1963)
  1186--1194.
\newblock \href {http://dx.doi.org/http://dx.doi.org/10.1063/1.1734411}
  {\path{doi:http://dx.doi.org/10.1063/1.1734411}}.

\bibitem{OliverKnoll1968}
D.~Oliver, G.~Knoll, Anisotropy of scintillation response of anthracene to
  neutron generated recoil protons and carbon ions, Nuclear Science, IEEE
  Transactions on 15~(3) (1968) 122--126.
\newblock \href {http://dx.doi.org/10.1109/TNS.1968.4324926}
  {\path{doi:10.1109/TNS.1968.4324926}}.

\bibitem{Tsukada1962286}
K.~Tsukada, S.~Kikuchi, Directional anisotropy in the characteristics of the
  organic-crystal scintillators, Nuclear Instruments and Methods 17~(3) (1962)
  286 -- 288.
\newblock \href
  {http://dx.doi.org/http://dx.doi.org/10.1016/0029-554X(62)90007-1}
  {\path{doi:http://dx.doi.org/10.1016/0029-554X(62)90007-1}}.

\bibitem{lbnlmuon} \textit{Atomic and Nuclear Properties of Materials: Anthracene.} [Online] Available: http://pdg.lbl.gov/2015/AtomicNuclearProperties/ [2015, Aug. 8]. Lawrence Berkeley National Laboratory, Berkeley, CA. 

\bibitem{nist} Berger, M.J., Coursey, J.S., Zucker, M.A., and Chang, J. (2005), \textit{ESTAR, PSTAR, and ASTAR: Computer Programs for Calculating Stopping-Power and Range Tables for Electrons, Protons, and Helium Ions} (version 1.2.3). [Online] Available: http://physics.nist.gov/Star [2015, Aug. 8]. National Institute of Standards and Technology, Gaithersburg, MD.



\end{thebibliography}

\end{document}